# The contribution of plastic sink-in to the static friction of single asperity microscopic contacts

Owen Brazil[1], John B. Pethica[2], George M. Pharr[1]

[1]*Department of Materials Science & Engineering, Texas A&M University, College Station, TX 77840, USA*

[2]*School of Physics, CRANN & AMBER, Trinity College Dublin, Ireland*

*Correspondence to: brazilo@tcd.ie

# Abstract

We report microscale friction experiments for diamond/metal and diamond/silica contacts under gigapascal contact pressures. Using a new nanoprobe technique which has sufficient dynamic range of force and stiffness, we demonstrate the processes involved in the transition from purely interface sliding at the nanoscale to the situation where at least one of the sliding bodies undergoes some plastic deformation. For sliding of micrometer-scale diamond spherical tips on metallic substrates, additional local plastic yielding of the substrate resulting from tangential tractions causes the tip to sink into the surface, increasing the contact area in the direction of loading and resulting in a static friction coefficient higher than the kinetic during ploughing. This sink-in is largely absent in fused silica, and no friction drop is observed, along with lower friction in general. The transition from sinking in within the static friction regime to ploughing in the sliding friction regime is mediated by failure of the contact interface, indicated by a sharp increase in energy dissipation. At lower contact pressures, the elastic interfacial sliding behaviour characteristic of scanning probe or surface force apparatus experiments is recovered, bridging the gap between the exotic realm of nanotribology and plasticity dominated macroscale friction. We delineate the material and geometric factors which determine the transition. We also find that some unexpected light is cast on the origins of the difference between static and dynamic friction.

# Main Text

## 1. Introduction

In classic literature on the sliding of solids at the macroscale,[1]–[3] plastic yielding and thus hardness plays a crucial role in our understanding of the origins of friction. It is directly involved in the relation between true contact area and load on real surfaces, which in turn explains general friction coefficients. It strongly affects wear processes, including in creating the run-in state,[4] and it provides a quantitative mechanism of energy dissipation in friction.[5] The more recent development of nanoscale mechanical probes has enabled investigation of contacts similar in scale to the individual asperities which underly macroscopic sliding friction.[6], [7] This has elucidated the mechanisms of energy dissipation which occur directly at the interface or in thin film lubricants and may not necessarily involve wear.[8], [9] However, until now, limitations of both the apparatus and the scale of atomistic simulations has meant that almost all nanotribology studies involve planar or near-planar interfaces and idealised geometries which are also largely unmodified by the sliding process. Generally speaking, such studies do not address plastic deformation mechanisms such as dislocation motion, albeit with some exceptions.[10], [11] It is therefore currently difficult to connect insights from nanotribology studies to friction at the larger scale and to the plasticity involved in a great many applications.

Laboratory single asperity contacts are often taken as representative, if somewhat idealised, analogues for the points of contact on macroscopic mating surfaces, where due to roughness, the true contact area is limited to a miniscule fraction of the apparent contact area.[12] Experimental issues such as low support frame stiffness and inter-axial coupling in surface force apparatus and scanning probe microscopy studies[6], [13] have made it extremely challenging to achieve in these experiments the high normal contact pressures and plastic deformation common to asperities on real engineering surfaces. Uncertainties in tip shape and contact area in the latter technique further cloud the picture.[14] As such, the role of plasticity in small scale friction and how it interacts with interfacial phenomena such as surface corrugation or adsorbed surface layers is not well understood. This is particularly true of the early stages of lateral loading when the contact transitions from static to kinetic friction, with most studies focusing rather on the less transient steady ploughing/grooving regime for simplicity.[10], [15], [16] Further, a sharp distinction has been often been drawn between the interface or surface dominated friction and wear and plasticity dominated friction which may not be reflective of the true behaviour of microscopic contacts, especially in the presence of the surface adsorbed layers found in real contacts outside ultra-high vacuum. [17]

Across a broad range of length scales and material types it is observed that the force required to initiate relative sliding between two bodies in contact is greater than the force required to maintain it, i.e., the static coefficient of friction $\mu_s$ is greater than the kinetic coefficient $\mu_k$.[1], [3], [18] At the microscopic, single asperity level, the lateral force needed to shear a junction and initiate sliding can be written in terms of the contact area of the junction $A_c$ and an interfacial shear strength, $\tau_0$, via: $Q_{shear} = \tau_0 A_c$.[1], [7], [9] Within this framework, the friction drop from $\mu_s$ to $\mu_k$ is often interpreted in terms of frictional ageing: as the two bodies are held in contact prior to lateral loading, chemical bonding may occur across the interface which, along with molecules interlocking and taking new equilibrium positions, raises $\tau_0$ and increases the quality or shear strength of the contact.[19], [20] Independently, time-dependent plasticity and creep may increase $A_c$ over longer time periods, thereby raising $Q_{shear}$ and $\mu_s$ above their values during sliding.[21], [22]

In this work, we use a newly developed dual-axis load-controlled nanoindenter[23] to apply GPa normal pressures to single asperity contacts with contact radii $a_c$ ranging from several hundred to several thousand nanometres, i.e., contacts that are about a micrometer in dimension. Several interesting observations are made on the nature of static friction for these plastically deforming contacts. First, we note the static friction maximum for diamond/ductile metal contacts coincides with significant sinking in of the tip into the substrate due to additional local plastic deformation during sliding. This sink-in is greatly reduced in more plastically-resistant fused silica substrates, in which no static friction maximum is observed, indicating local plastic deformation and sink-in is largely responsible for the friction drop from $\mu_s$ to $\mu_k$ in more ductile materials. We examine this plastic sinking in of the tip under two frameworks: first Tabor's junction growth model[24], [25] for purely ductile materials wherein the addition of lateral load produces a multi-axial stress state that causes further plastic yielding, and second, for higher yield strain materials, there is an asymmetry of stress across the contact due to the more significant elastic component of strain.[26] This leads to stress concentration and hence deformation in the direction of the lateral load on the front side of the sliding tip and a reduction at the rear.[26], [27] We find both models have applicable regions, their extent depending on the contact geometry and elastic modulus to hardness ratio of the substrate. We show that the transition from static to kinetic friction for plastic contacts coincides with the tip ceasing to sink in and beginning to rise out of the substrate, before achieving an essentially constant depth during steady ploughing and grooving. A rapid increase in dissipative energy at this point suggests the transition is facilitated by failure of an interfacial layer. Finally, we show that at sufficiently low normal loads, our experiments re-capture the essential physics of traditional wearless single asperity studies, indicating that the elastic and plastic friction domains are not as separate as is sometimes assumed.

## 2. Materials and Methods

Lateral loading friction experiments were carried out using the Gemini (Nanomechanics Inc, TN, USA) dual axis nanoindentation system. The instrument consists of two independent i-Micro load actuators and is capable of applying loads of up to 50 mN simultaneously in the normal and lateral directions. The actuators are connected to the indenter tip by means of two glass slides of approximately 1.5 cm length and 0.5 cm width that join at a special tip mounting block. The lateral frame stiffness of the instrument was measured as 105,000 N/m, as was described previously,[23] and is treated as a one dimensional spring in series with the contact so as to obtain true displacement of the tip through the substrate. The vertical frame stiffness is approximately $10^6$ N/m. All initial indents were performed at constant $\frac{\dot{P}_z}{P_z} = 0.1\ s^{-1}$ to a prescribed normal load prior to lateral loading. This normal load was then held constant throughout the duration of lateral loading/unloading. In order to apply a lateral load $Q_x$ to the contact, a lateral force $F_x$ was applied via the horizontal load actuator at a constant rate of 0.25 mN/s. $F_x$ and the resultant $Q_x$ are related at any time point via $Q_x = F_x - k_x(d_x - d_0)$, where $k_x$ is the stiffness of the column of the lateral actuator, equal to approximately 540 N/m, $d_x$ is the present lateral position of the tip as measured via the lateral capacitance gauge, and $d_0$ is the lateral position immediately prior to the application of $F_x$. Within the static friction domain where lateral displacements are small, $F_x$ and $Q_x$ and their loading rates $\dot{F}_x$ and $\dot{Q}_x$ are approximately equivalent; however, in the kinetic friction regime, significant divergence emerges as $d_x$ grows large and the stiffness of the contact becomes low enough to be comparable to $k_x$. Contact stiffnesses in both directions are measured via the continuous stiffness method (CSM),[28] with each actuator equipped with a dedicated lock-in amplifier.

The fused silica and polycrystalline aluminium samples used in this work were supplied as reference materials accompanying the nanoindenter via Nanomechanics Inc. The single crystal nickel sample was purchased from Surface Preparation Laboratory, Zandaam, NL and mechanically polished to a mean roughness of less than 5 nm. All three indenter tips used in this work were supplied by Micro-Star Technologies, Huntsville, TX, USA.

## 3. Results and Discussion

**3.1 Friction of single asperity plastic contacts.**

Fig. 1a shows the essential contact geometry as modeled and experimentally studied here using the two axis nanoindentation system.[23] A normal load $P_z$ on the order of a few mN is applied to a 1 micron radius sphero-conical diamond tip, sufficient to indent the substrate and produce significant plastic deformation in soft metals. $P_z$ is then held constant throughout the remainder of the experiment. A lateral force is then applied via a second actuator connected to the tip and oriented orthogonal to the surface normal, resulting in a lateral load $Q_x$ applied to the contact which continuously increases at $\dot{Q}_x \sim 0.25$ mN/s prior to kinetic sliding/ploughing (see methods for full discussion). The vertical and lateral displacements of the tip into the surface, $\delta_z$ & $\delta_x$, are monitored continuously, while the vertical and lateral contact stiffnesses, $S_z$ & $S_x$, are measured by superimposing small dynamic oscillatory loads onto the quasi-static loads via the continuous stiffness method commonly employed in nanoindentation.[28], [29] These dynamic loads are typically 2% of the magnitude of the quasi-static load signal, that is, sufficiently small so as to introduce no significant change in the overall contact mechanics.[30], [31] From $S_z$ and known elastic constants, the contact area $A_c$ and radius $a_c$ can be inferred during normal loading.[32] Typical radii prior to applying $Q_x$ are $a_c \sim 500$ nm for the contacts considered here. In Fig. 1b, we plot $\delta_x$ versus the friction coefficient $\mu = Q_x/P_z$ for three substrate materials: a single crystal nickel <111> surface (ScNi), polycrystalline aluminium (PcAl), and fused silica (FS). All samples are loaded normally to an initial depth of $\delta_{z_0} = 200$ nm, corresponding to mean contact pressures of 2.72, 0.48, and 10 GPa, respectively. We note the diamond tip is sufficiently hard when compared to the sample materials that we may assume all plastic deformation is limited to the substrate.[33] Analysing the friction curves, we see the diamond/metal contacts are distinguished from the diamond/silica contact not only by significantly higher $\mu$ throughout the experiment, but also by a $\mu_s$ peak that is higher than the $\mu_k$ during ploughing. Examination of the tip depth $\delta_z$ during lateral loading in Fig. 1c shows that for the metals, significant sink-in occurs in the "static" friction portion of the curves prior to ploughing, with $\delta_z$ increasing by approximately 100 nm for PcAl and 120 nm for ScNi. That the maximum in $\delta_z$ in Fig. 1c occurs concurrently with the $\mu$ maximum in Fig. 1b suggests this tip sink-in is strongly influencing the $\mu_s$ peak in the metals. Conversely, for the diamond/silica contact, the sink-in is much smaller, around 20 nm or 10% of the initial tip depth, and no $\mu$ maximum can be identified in Fig. 1b, i.e., $\mu_s = \mu_k$. In Fig. 1d, we displace a ScNi contact by 200 nm laterally and then unload, showing the displacement preceding the $\mu_s$ maximum is permanent, inelastic, and irrecoverable. Fig. 1e shows the accompanying sink-in is similarly irrecoverable, while Figs. 1f & 1g repeat the experiment for the diamond/fused silica contact, with the 200 nm lateral displacement producing a much smaller sink-in of ~ 10 nm, in keeping with the results of Fig. 1c.

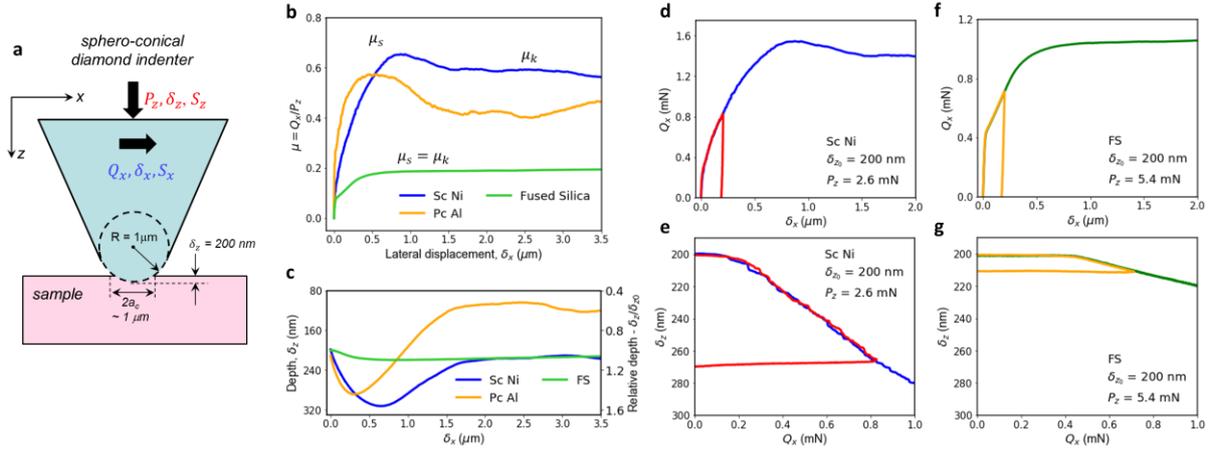

**Figure 1. Friction behaviour of plastic single asperity contacts** (a) Sketch of the contact geometry considered here. (b) Lateral friction curves ($\delta_x$ - $\mu$) curves into the single crystal nickel, polycrystalline aluminium, and amorphous fused silica samples probed here. (c) Vertical tip sink-in depth $\delta_z$ as a function of the lateral displacement $\delta_x$. The maxima in $\delta_z$ match well with the maxima in $\mu$ in (b), indicating junction growth plays a significant role in static friction. (d) Lateral load $Q_x$ versus displacement $\delta_x$ curves for the nickel sample. The red curve loaded to $\delta_x = 200$ nm shows lateral displacement in the 'static' friction regime is irreversible. (e) Vertical sink-in of the tip for the same experiment. Again, a large irreversible deformation is observed. (f), (g) Equivalent curves for fused silica.

It is our assertion that the irrecoverable sinking-in of the indenter tip under combined loading conditions is responsible for the high static friction seen in the diamond/metal contacts, with the amount of sink-in determining how high $\mu_s$ becomes. The addition of the lateral load $Q_x$ raises the total shear stress within the contact, and, given that our substrate materials are already deforming plastically under the normal load alone, must result in further yielding, which manifests as tip sink-in. For the harder fused silica substrate, the low elastic modulus to hardness ratio $E/H \sim 8$ limits the amount of ductile yielding and additional sinking-in that occurs, resulting in lower friction than the metals, where $E/H \sim 100$. Combined with additional pile-up at the front of the tip, this sink-in leads to an increase in the contact area at the front of the tip and more substrate material to displace, resulting in greater resistance to sliding and a higher coefficient of friction.

To verify this hypothesis, in Fig. 2 we examine the residual deformation left in the substrate at various stages of the friction curve for the diamond/ScNi contact. Fig. 2a replots the original friction curve seen in Fig. 1b, along with three separate indentation/lateral loading experiments where the nickel is laterally loaded to $Q_x$ values lying within the static friction/sink-in regime, again initially under 2.72 GPa normal pressure. Test (i) considers a $Q_x$ small enough that the lateral response is essentially elastic, with minimal developed residual displacement. Test (ii) is loaded to a $Q_x$ where there is significant irrecoverable lateral displacement, and finally test (iii) is loaded to a $Q_x$ corresponding to the peak friction value at $\mu \sim 0.6$. Fig. 2b shows residual deformation maps of tests (i-iii) obtained via AFM, with the lateral loading direction oriented from left to right. As expected, test (i) is essentially unaffected by the small $Q_x$, with the residual deformation primarily the result of the normal load and pile-up distributed symmetrically around the contact impression in slip steps reflective of the <111> surface orientation. For test (ii), significant pile-up is present in the direction of $Q_x$, indicative of additional plastic deformation. The size of residual contact impression also increases in the direction of loading, displaying a distinct asymmetry. AFM depth profiles extracted from Fig. 2b and plotted in Fig. 2c further emphasize this point. Sink-in continues to increase for test (iii) in Fig. 2b, along with more pile-up forming at the front edge of the contact. Comparing tests (i) and (iii), the total interior surface area of the residual contact impression has grown from 1.12 μm² to 2.90 μm², an increase of approximately 160%. An image of the residual deformation for a contact loaded much further still into the kinetic friction/ploughing domain is shown in Fig. 2d, where the rise in tip depth to a shallower, essentially constant $\delta_z$ during ploughing implied by Fig. 1c is verified. The cross-sectional width of the scratch track decreases during ploughing, reflective of decreasing $A_c$. We also note that the residual depth profile of the contact measured by AFM and the tip depth $\delta_z$ reported in real time by the indenter match very well as is shown in in Fig. 2e, indicating there is minimal elastic recovery following the removal of load and that the displacements reported by the dual-axis indenter are representative of real displacements, both normal and lateral, at the contact.

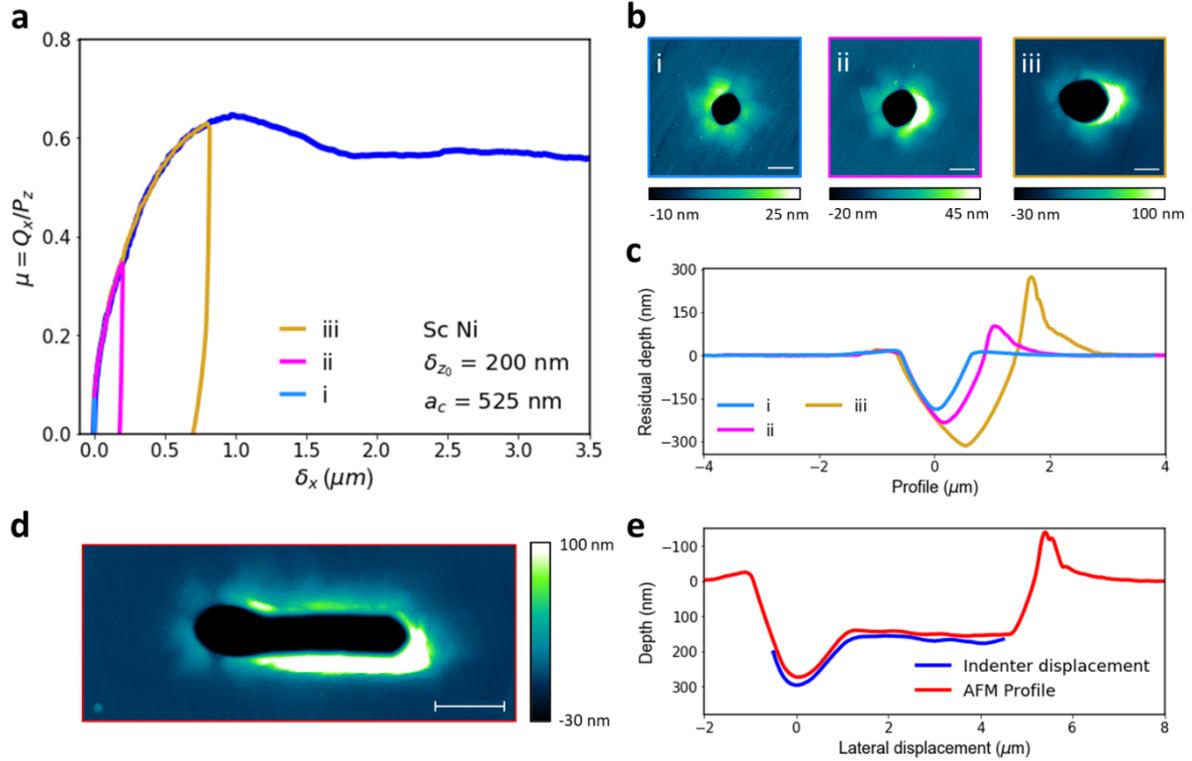

**Figure 2. Plastic sink-in in nickel** (a) Friction curves for single crystal nickel, with curves i-iii loaded to incrementally higher $Q_x$, but insufficient to cause ploughing. (b) Corresponding AFM residual topography maps for indents i-iii. The lateral load is applied from left to right in the image. Scale bars are 1 μm. (c) Accompanying AFM height profiles taken from left to right across the centre of the images. (d) AFM topography map for a large scratch. 2 μm scale bar. (e) Comparison of the residual depth profile from (d) with the tip depth measured in-situ during lateral loading by the indenter.

### 3.2 Plastic junction growth during static friction.

In the introduction we have mentioned two possible mechanisms that contribute to the added sinking in of the tip into the substrate under combined loading, which represent two classic schools of thought regarding the influence of combined stresses on the mechanics of plastic contacts. First is the idea that by adding a lateral load $Q_x$ to the contact, the total shear stress in the contact increases. Assuming a sharp, fixed plastic yield stress $Y_0$ and in the absence of significant strain hardening, increasing the total shear stress at the contact must result in an increase in the contact area and mutual approach of the contacting bodies (sink-in.) Following a series of macroscopic friction experiments conducted throughout the 1950's by researchers such as Courtney-Pratt and Eisner,[34], [35] this theory of plastic junction growth was put on rigorous footing towards the end of the decade by Tabor, who derived a relationship between the applied loads and the contact area $A_c$ of the form [24]:

$$\frac{1}{\alpha}\left(\frac{Q_x}{P_z}\right)^2 = \left(\frac{A_c}{A_{c_0}}\right)^2 - 1, \qquad (2)$$

where $A_{c_0}$ is the contact area resulting from conventional indentation before applying the lateral load $Q_x$ and $\alpha$ is a material and geometry dependent constant, typically on the order of 10 for 3D contacts. Because we are able to accurately measure $P_z$, $Q_x$, and the normal contact stiffness $S_z$ directly, the two-dimensional indentation system we employ allows us to study plastic friction theories such as this junction growth model for the first time at the single asperity scale for contact radii of several hundred nanometers. We test the applicability of the junction growth formulation at the microscale by assuming that $A_c$ is proportional to $S_z^2$, as is done in conventional CSM indentation analysis,[32] so that equation (2) becomes:

$$\frac{1}{\alpha}\left(\frac{Q_x}{P_z}\right)^2 = \left(\frac{S_z}{S_{z_0}}\right)^4 - 1, \qquad (3)$$

where $S_{z_0}$ is the initial normal contact stiffness. Figure 3(a) plots the right-hand side of equation (3) versus $\left(\frac{Q_x}{P_z}\right)^2$ for two contacts: the red curve represents the response of a PcAl substrate indented with a diamond Berkovich three-sided pyramidal tip with an equivalent cone angle of 70.3° under a normal load of $P_z = 40$ mN and then laterally loaded. The blue curve represents the same Berkovich tip loaded into fused silica to the same $P_z$. The Berkovich tip offers a distinct advantage over the sphero-conical geometry previously employed in that the tip is self-similar, meaning the state of strain experienced by the substrate should be independent of normal load/depth prior to lateral loading. As a consequence of this, the Berkovich tip should result in plastic deformation of the substrate even at the lowest loads, so long as tip rounding is not too severe. Returning to Fig. 3a, for both contacts the lateral load is applied in a pyramid 'edge forward' fashion, rather than a pyramid 'face forward' orientation. The effective contact radius $a_{c_0} = \sqrt{A_{c_0}/\pi}$ prior to lateral loading is 2580 nm for PcAl and 680 nm for FS. From Fig. 3a, we see that the predicted linear relation between the two quantities suggested by equations 2 and 3 is indeed observed prior to ploughing, validating Tabor's model for single asperity contacts. The measured values for the proportionality constant of $\alpha_{PcAl} = 8.5$ and $\alpha_{FS} = 4.2$ are reasonable in the context of values observed for macroscopic contact pairs,[24], [36], [37] with the higher value for the ductile aluminium substrate suggesting higher $\alpha$ corresponds to larger junction growth and, in our case, greater sink-in of the diamond tip. We note the second linear region, beginning at about $(Q_x/P_z)^2 \sim 0.10$ for PcAl, deviates from Tabor's solution and corresponds to the tip rising out of the substrate, as was observed in Fig. 1c. The dramatic drop in stiffness following this portion of the curve then corresponds to ploughing during kinetic friction. In Figs. 3b and 3c, we extend the experiment, varying the normal load over a range of 5 – 30 mN, corresponding to $a_{c_0}$ values of 2815 - 6950 nm for PcAl in 3b and 430 - 1295 nm for FS in 3c. The value of $\alpha$ for both materials appears essentially constant with contact size and normal load down to a few hundred nm, indicating that in this geometry junction growth holds as an accurate descriptor for the early stages of sheared contacts where the interface is composed of several thousand atoms or fewer. However, at shallower depths (smaller $a_c$) the departure from the junction growth model and the transition to ploughing occur sooner, suggesting a reduced role for plasticity in the friction of very small contacts when compared to interfacial contributions. We consider this transition further in section 3.3.

We note that in Figs. 3a and 3c for the fused silica contact, there is a small additional linear region that precedes which is well described by junction growth and appears reasonably constant in size at all $a_c$. This region corresponds to the initial, low load region of the fused silica lateral load versus displacement curve studied in Fig. 1f, for which there is minimal tip sink-in, as was shown in Fig. 1g. The origins of this seemingly elastic region whose existence is contrary to interpretations of plasticity that feature a fixed yield stress $Y_0$ such as those we have employed here are not entirely clear at present; however we speculate that is arises from a combination of instrumentation effects such as tip rotation upon the immediate application of $Q_x$ and interfacial friction which may screen the bulk material from some of the added shear stress.[13] This region shall be the subject of a forthcoming publication.

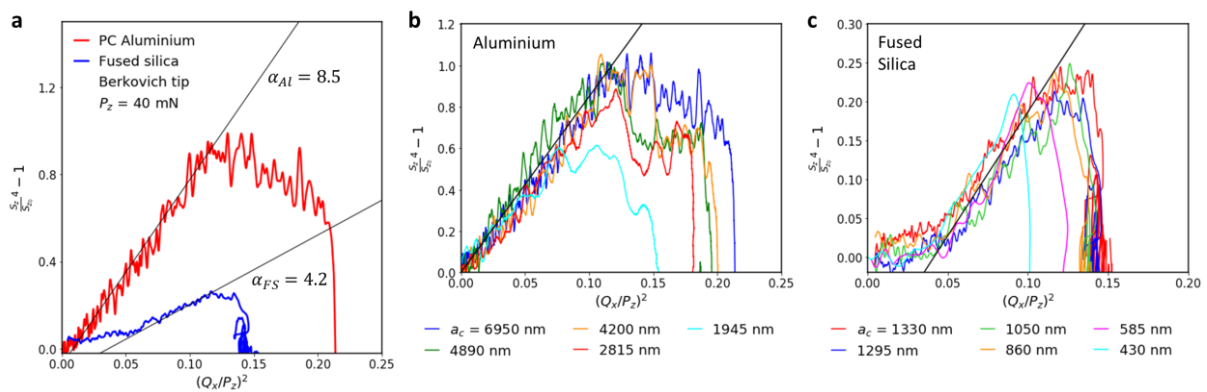

**Figure 3. Modelling plastic sink-in as junction growth.** (a) Equation 3 plotted for plastically deforming fused silica (blue) and aluminium (red) substrates indented and laterally loaded with a diamond Berkovich pyramidal indenter tip. The predicted linear slope indicates good agreement with Tabor's model. (b) Junction growth curves for aluminium as a function of initial $a_c$. The slope appears constant with contact size and normal load, likely a result of the tip self-similarity. (c) The same experiment repeated for fused silica, again showing a constant slope with $a_c$.

While the junction growth model outlined above initially works well for the sharp, deep indentation geometry imposed by the Berkovich tip in Fig. 3, we find that it fails utterly in describing the behaviour of ductile substrate materials in the smoother and flatter contact geometry encountered when using the sphero-conical tip at low contact depths. In Fig. 4a we plot $\left(\frac{Q_x}{P_z}\right)^2$ versus $\left(\frac{S_z}{S_{z_0}}\right)^4 - 1$ for the contacts formed of the 1 μm radius tip and the PcAl (red) and FS (blue) substrates at an initial contact depth of $\delta_{z_0} = 200$ nm, as were studied in Fig. 1. While for the FS contact there is similarity to the junction growth behaviour observed with the Berkovich tip, as shown by the linear positive slip in the inset in Fig. 4a, for the PcAl contact the behaviour is markedly different. Instead of linearly increasing, a sharp and dramatic decrease in $\left(\frac{S_z}{S_{z_0}}\right)^4 - 1$ occurs at $\left(\frac{Q_x}{P_z}\right)^2 \sim 0.02$, and prior to this point the slope is essentially flat. We note from the position of the orange ϕ symbol in Fig. 4a and the lateral loading curve in Fig. 4b that this sudden $\left(\frac{S_z}{S_{z_0}}\right)^4$ drop occurs within the static friction limit, well before the transition to ploughing, which itself is denoted in 4a & 4b by the green asterisks. To explain this behaviour, we return to the second mechanism that occurs to cause tip sink-in when a lateral load is introduced to the contact; adding $Q_x$ causes the direction of the total load vector applied to the contact to shift towards the direction of lateral loading. As such, material at the rear of the tip is unloaded, and the total contact area actually reduces. The substrate material located at the front of the tip now carries a much higher total load, which causes further yield and therefore sink-in. This hypothesis was put on sound theoretical footing by a series of slip line field theory studies, most notably by Johnson, for two-dimensional rigid wedges first indented and then laterally loaded into perfectly plastic substrates.[26] These studies explained sink-in via increased stress at the front of the slider rather than an increase in the total stress across the entirety of the contact and agree well with the experimental findings of Challen and Oxley, who studied a macroscopic slider loaded into a ductile aluminium substrate in a plane-strain geometry.[27] Within this formulation we may readily explain the behaviour seen in Fig. 4a by considering the normal stiffness of the contact, $S_z$, as $Q_x$ is increased, which is plotted in Fig. 4c for PcAl. Following an initial region where the stiffness is essentially constant, at $Q_x = 0.1$ mN $S_z$ drops by over half its initial value from 95,000 N/m to just over 40,000 N/m. This stiffness drop corresponds to the reduction in contact area as the back side of the tip is unloaded. $S_z$ then begins to gradually increase from 40,000 N/m to 60,000 N/m at $Q_x = 0.35$ mN. This increase is due to the tip sinking in as material toward the front of the tip yields and the contact area again begins to increase before steady ploughing is established.

The behaviour of the FS contact can also be explained with this model. As is seen in Fig. 4d, the lateral stiffness $S_z$ increases for FS during lateral loading, giving a behaviour in Fig. 4a that resembles the junction growth behaviour of Fig. 3. Upon applying $Q_x$, the rear side of the tip is unloaded as in the case of PcAl. However, the low $E/H$ ratio allows for more elastic recovery of substrate material at this side of the tip, which is absent in the more plastic aluminium. This elastic recovery keeps material at the rear in contact in the tip, meaning the total contact area, and therefore stiffness, increases, despite less plastic yielding and sinking-in occurring.[38] The different behaviours of a high and low $E/H$ material in this geometry are outlined schematically in Fig. 4e, where the dashed red line schematically indicates the position of the tip prior to applying $Q_x$. In considering the two mechanical models presented in Figs. 3 and 4 and when they are accurate descriptors of plastic static friction, we would suggest they both represent limiting cases. Junction growth is an applicable model in the case of sharp, deep indentation geometries where the tip is well embedded in the substrate and therefore more constrained by material around it. In this scenario the contact approximates the welded plastic junctions studied by Tabor and co-workers.[24], [36] For flatter, more shallow contacts however, where tip is less embedded into the surface, the stress concentration model is more appropriate, depending on the substrate material properties. The 2D slip line field formulation makes some account for this fact by incorporating a $tan(\vartheta)$ dependency for the cessation of sink-in, where $\vartheta$ is the included angle of the wedge. As such, sharper wedges (small $\vartheta$) are subject to more sink-in than blunter ones, generally in keeping with our findings. Given both models were developed for a perfectly plastic constitutive response, it is not surprising that for real contacts, deviation from both emerges as a result not only of geometric factors, but also due to the elastic properties of the substrate.

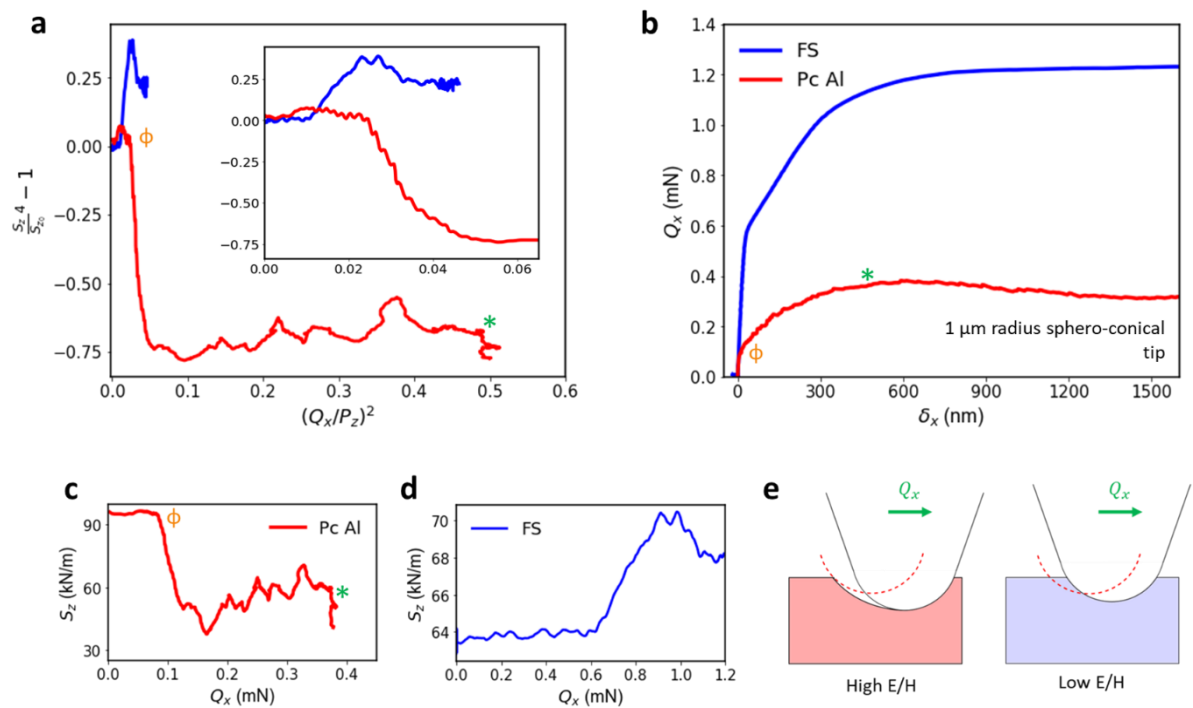

**Figure 4. Failure of junction growth model for sphero-conical tips.** (a) Equation 3 plotted for FS and PcAl contacts indented to $\delta_{z_0} = 200\ nm$ with the 1 μm radius sphero-conical tip. Inset shows zoom-in of early loading behaviour. (b) Lateral load versus displacement curves for PcAl and FS showing positions of interest relating to (a). (c) Normal contact stiffness $S_z$ as a function of lateral load for PcAl contact. (d) Equivalent curve for FS. (e) Schematic of how elastic recovery affects total contact area, and therefore $S_z$, for high and low $E/H$ substrates.

The results of Fig. 4 suggest that for plastic single asperity contacts, friction is determined not by the total area of contact, as implied by equation 1, but rather the contact area at the front of the tip in the direction of sliding. While this line of thinking is commonly employed to describe the behaviour of ploughing contacts in nano-scratch testing,[15], [39] it has not generally been applied to the early, transient stages of multi-axial loading. This is likely a result of the fact that conventional nano-scratch apparatuses operate in a displacement-controlled manner, making detailed study of the static friction regime rather challenging.[40] To confirm this interpretation and verify that the increase in $S_z$ (and therefore $A_c$) for the FS contact but not for PcAl in Fig. 4 arises from elastic recovery due to its lower $E/H$, we conduct a series of finite element simulations of a rigid 2D cylindrical slider of 1 μm radius indented and then laterally loaded into elastic – perfectly plastic half-spaces. Using the ABAQUS 2019 Explicit FEA package (Dassault Systemès), we indent the slider to a depth of $\delta_{z_0} = 200$ nm and, as in our experiments, then hold the normal load $P_z$ constant during the remainder of the test. The slider is then displaced laterally by 1.5 μm at a constant velocity of 0.6 μm/s. This method is chosen over applying a prescribed lateral loading rate $\dot{Q}_x$ as in our experiment due to greater numerical stability. An elastic modulus of 10 GPa and a Poisson's ratio of 0.3 is used for the half-space in all simulations, while plastic yield is modelled using a Von Mises yield stress $Y_0$. Three values of $Y_0$ are used; $Y_0 = 0.6$, 0.3, and 0.15 GPa. Assuming hardness to be related to the yield stress via Tabor's relation, $H \sim CY$, where C is a constant approximately equal to 3 for ductile metals,[26] these values give $E/H$ ratios of 5.56, 11.1, and 22.2 respectively. These values are quite low and are more representative of the fused silica contact rather than the metal but were chosen as they generated greater simulation stability than lower values of $Y_0$. The surface interaction between the slider and half-space is specified to be frictionless, so that all resistance to sliding originates from substrate deformation rather than interfacial effects. Fig. 5a shows a close-up of the contact geometry (the half-space is much larger than shown here, see methods) at the early stages of lateral displacement for the $E/H = 11.1$ simulation. The colour scheme represents von Mises stress contours. Fig. 5b shows the same simulation at a more developed stage, where it is observed that the depth of the slider into the substrate had increased. Fig. 5c plots the measured friction force $\mu$, taken to be the force required to displace the slider laterally divided by the constant normal load, as a function of $\delta_x$. As expected, Fig. 5c shows that more ductile, higher $E/H$ materials experience greater friction. Fig. 5d plots the increase in slider depth during lateral displacement, with the more ductile materials subject to greater plastic sink-in, in accordance with the experimental findings of Figures 1 and 2. In Fig. 5e, we plot the normalised contact area $A_c/A_{c_0}$ for the three half-spaces and observe that in all three cases there is an increase in contact area at very small $\delta_x$, likely due to additional plastic yielding via the junction growth mechanism. For the $E/H = 22.2$ and $E/H = 11.1$ materials, this increase is small, approximately 1% of the initial $A_c$, and quickly subsides as the rear of the slider is unloaded. For the less ductile $E/H = 5.56$ material, however, despite significantly less plastic yield as evidenced by less sink-in in Fig. 5d, the $A_c$ increase is much larger at approximately 6% of the initial value and persists much longer. The validates our interpretation of the experimental findings of Fig. 4, where the disparity in the stiffness behaviour of the FS and PcAl contacts is attributed to the vastly different $E/H$ of the two materials. Further, these findings show that total contact area is not necessarily as good an indicator of the friction resisting sliding of a plastically deforming microcontact as it is for an elastically deforming one, since material in the path of the tip is more important than material towards the rear. Given the experimental findings of Figure 1 and the simulations of Figure 5, we suggest tip sink-in to be a more quantitative descriptor.

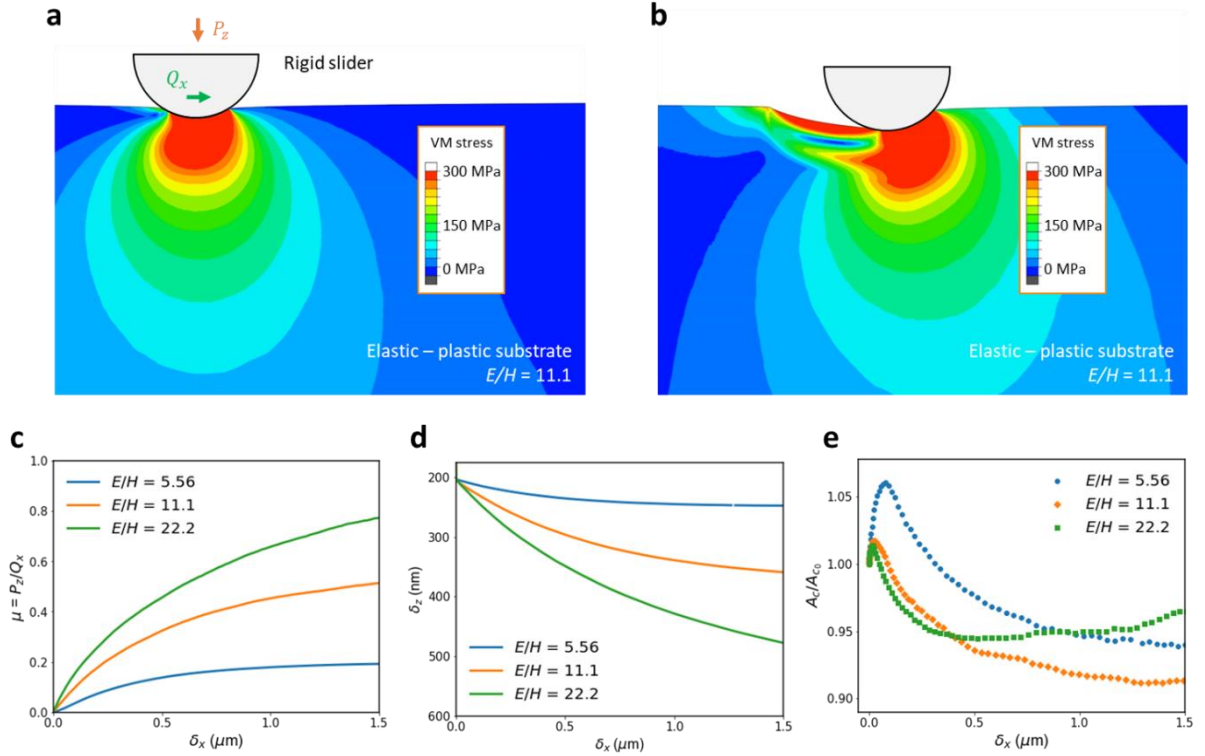

**Figure 5. Finite element simulations of lateral loading experiments.** (a) Schematic of the 1 μm radius two-dimensional rigid slider indented and laterally loaded into an elastic – plastic half space with $E/H$ = 11.1. Colour contours represent the Mises stress. (b) The same simulation at a greater lateral displacement. (c) Simulation friction curves for three $E/H$ values. (d) Tip sink-in as a function of lateral displacement. (e) Total contact area between slider and substrate during lateral displacement. Despite less plastic sink-in, the lowest $E/H$ material sees the largest increase in $A_c$ due to elastic recovery at the tip rear.

### 3.3 Interface slip and the transition to ploughing.

To this point we have described how sink-in of the tip leads to a greater static coefficient of friction than kinetic. However, the origins of the transition between the static domain where the depth of the tip into the substrate increases and the kinetic steady ploughing domain where the tip depth is essentially constant is not yet clear. In the slip line field theory formulation, the 2D wedge continues to sink into the surface until the pressure on the rear face of the tip reaches zero (i.e., the rear face is totally unloaded), at which point the wedge begins to ride up out of the substrate.[26], [27] In this highly simplified model, this occurs when the lateral load becomes $Q_x = P_z \tan(\vartheta)$. However, this implies a transition between static and kinetic coefficient states with no friction drop, i.e. $\mu_s = \mu_k$, which is at odds with our observations in nickel and aluminium. When formulating his junction growth model, Tabor suggested the transition from local yield to ploughing was in practice mediated by shear failure of an interfacial layer composed of third body contaminants such as oxide layers, adsorbed hydrocarbons, and water menisci.[24] This idea was initially proposed to explain why experiments conducted under high vacuum exhibited much more junction growth than those conducted in ambient conditions, with values of $\mu \sim 100$ routinely recorded in the former. The aforementioned experiments of Challen and Oxley also highlight the importance of third body lubricants and adsorbed materials, with a large friction drop ($\mu_s > \mu_k$) observed in the cases of dry or poorly lubricated contacts, whereas $\mu_s \sim \mu_k$ in the presence of a well-matched boundary lubricant.[27] Tabor's concept is supported by modern molecular dynamics studies of single asperity friction showing that static friction between incommensurate lattices emerges only when a layer of hydrocarbons is present between the two contact bodies.[9] The experiments of Homola also make clear the critical role of thin interfacial layers,[17] as does more recent work of Li et al. on the friction of layered two-dimensional materials.[41] Further, such layers can affect mechanical properties of the deformed volume such as hardness.[42] Should the failure of interfacial layers enable the transition from static to kinetic friction in elastically deforming systems, and given they are observed to be important in plastic wedges[27], it is eminently reasonable to conclude they may play a similar role in mediating the transition from sink-in to ploughing in the plastic contacts we study here. To explore this, in Figure 6 we compare the tip depth $\delta_z$ during lateral loading to the lateral CSM lock-in amplifier phase angle, $\theta_x$. In both nanoindentation and atomic force microscopy, the phase angle is a powerful indicator of energy dissipation, increasing significantly as more and more energy is lost to dissipative mechanisms, and is consequently used to track dissipative processes in adhesion, plasticity, and slip.[23], [31], [43] Fig. 6a plots $\delta_z$ and $\theta_x$ versus lateral displacement $\delta_x$ for the ScNi <111> contact studied in Figures 1 and 2. During the portion of Fig. 6a where the tip sinks in, $\theta_x$ increases steadily, most likely the result of the additional plasticity generated by $Q_x$. At $\delta_x$ = 0.75 µm, the slope of $\theta_x$ abruptly increases, indicative of the commencement of an additional energy dissipating process. The phase angle increase coincides with the point at which the tip begins to rise out of the surface. This indicates that the transition from sink-in to ploughing is mediated by initiation of a secondary dissipative process, which is likely to be the interface failure and relative sliding of interface molecules. Fig. 6b presents similar data for the fused silica contact. We note that in this case, $\theta_x$ has a reasonably constant non-zero value before the rapid increase at interface failure, likely a result of less additional plasticity under shear loading than the more ductile nickel. In other respects, however, the behaviours of the two materials are similar.

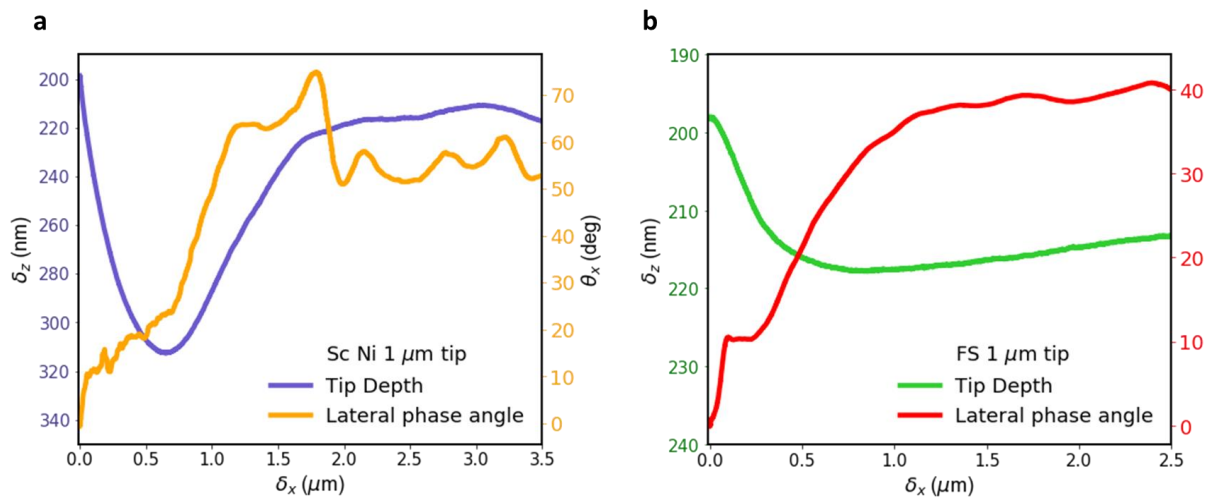

**Figure 6. Phase angle behaviour during lateral loading.** (a) Plots of lateral CSM phase angle $\theta_x$ (yellow) and tip depth into the surface (purple) for the ScNi contact studied in Figure 1. The transition from tip sink-in to rising out corresponds with a dramatic increase in $\theta_x$, indicating more energy dissipation. (b) Equivalent curves for the FS contact showing similar behaviour.

As a result of the large normal pressures we have imposed in our experiments, the friction of the contacts studied thus far in this paper in the earliest stages of lateral motion after indentation has largely been dominated by plastic deformation of the substrate. Under lower pressures, however, the elastic interfacial behaviour common to AFM and SFA experiments may be reproduced. We demonstrate this in Figure 7, where we additionally employ a 5 μm radius sphero-conical indenter, which enables us to probe lower contact pressures than either the Berkovich or 1 μm radius tips at equivalent normal loads. The three geometries are compared in Fig. 7a. Fig. 7b shows friction curves for the three tips indented and laterally loaded into FS under normal loads of 3 mN for the 5 μm tip, 5.6 mN for the 1 μm tip, and 50 mN for the Berkovich tip. These correspond to respective mean normal pressures of 4.25 GPa, 10 GPa, and 9.2 GPa. Fused silica was chosen for this experiment over the metals due to its higher hardness ensuring a larger and more experimentally accessible elastic region. Conventional load versus depth indentation curves are included in the supplement to show the 5 μm tip contact is elastic while the other two are plastic or elastic-plastic at these normal loads. Despite the differences in contact geometry, the shape of the 1 μm radius and Berkovich tips' friction curves are broadly similar, with a slope change at $\mu \sim 0.08$ corresponding to the initiation of sink-in and then a second slope change as the contact transitions to ploughing. However, in the case of the elastic contact formed with the 5 μm tip, the sink-in region of the curve is absent, with a single-step transition to sliding at $\mu \sim 0.08$. In Fig. 7c we confirm that no significant plasticity is generated during sliding for the contact formed with the 5 μm tip by plotting the depth of the tips $\delta_z$ normalized to the initial tip depth before lateral loading $\delta_{z_0}$ for the three contacts in 7c. No sinking-in is observed for the elastic contact, whereas for the plastic contacts a sink-in of 5 – 10% occurs. For the elastic contact, resistance to sliding arises from physical phenomena located at the contact. To shear, the interface adhesion and chemical bonding have to be overcome.[6], [13], [20] The striking result of Figs. 7b & 7c is that for the plastic contacts, these contributors to frictional resistance occur at approximately the same $\mu = {Q_x}/{P_z} \sim 0.08$. Not only does this imply these effects scale reasonably linearly with $P_z$ for the fused silica/diamond contacts considered, but also demonstrates the additional frictional resistance introduced by plastic deformation can be clearly identified over these interface effects, and its influence studied in isolation. In Fig. 7d we show the distinctive shape of the lateral loading curve for plastic contacts is a function of normal load $P_z$ rather than exclusively tip geometry by performing scratch tests with the 1 µm radius tip into FS under normal loads ranging from 0.25 – 6 mN, covering the elastic, elastic-plastic, and fully plastic regimes. Again, accompanying conventional nanoindentation load – displacement curves for the normal loads used in fig. 7d are included in the supplement so as to verify which contacts are elastic and plastic prior to lateral loading. Fig. 7d shows the second slope region associated with sink-in decreases in size at lower $P_z$ and that at the lowest $P_z = 0.25$ mN elastic behaviour similar to that encountered in fig. 7b with the larger tip is recovered. This transition from plastic or elastic-plastic scratching/ploughing at high $P_z$ to wearless sliding at lower loads is also captured in the lateral phase angle $\theta_x$, which is plotted in Fig. 7e for four of the experiments of Fig. 7d (same colour scheme.) For the high load scratching experiments, the same plateau associated with the sink-in regime is observed before the transition to sliding. However this region decreases in size as $P_z$ is reduced and is entirely absent from the $P_z = 0.25$ mN curve. In short, at low loads and a sufficiently near to flat contact geometry, sliding takes place directly upon application of lateral load. Increasing contact indentation causes an initial region of plastic deformation and lower energy dissipation; the transition to full sliding is increasingly delayed the larger the initial indent. We may say with some degree of confidence that the geometries and methods employed here to study plastically deforming single asperity contacts at lower loads and larger tip radii are able to replicate the elastic, largely wearless, interfacial sliding experiments typical of SFA and AFM studies. They therefore serve as an important bridge between the world of laboratory-based micro/nanotribology experiments and conventional plasticity-dominated engineering friction studies.

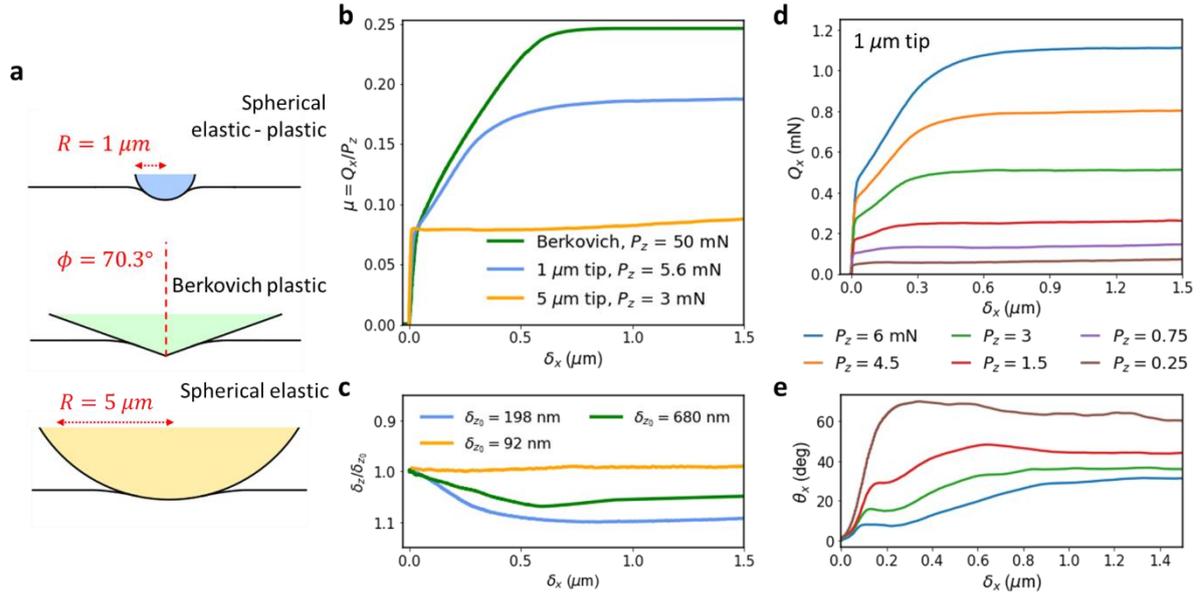

**Figure 7. Recovery of elastic interface sliding at low load.** (a) Contact geometries studied in this work. (b) Friction curves in FS for the 1 μm radius tip and Berkovich tip in the plastic limit, as well as the 5 μm radius tip in the elastic limit. (c) Relative tip depth as a function of lateral displacement for the three contacts. No sink-in is observed for the elastic contact. (d) Lateral load versus displacement curves taken with the 1 μm radius tip in FS at a series of different loads. The shape of the curve changes notably at lower loads, approaching that of the elastic curve in (b). (e) Lateral phase angle for 4 curves taken from (d). At lower loads, the $\delta_x$ plateau disappears, indicating the sink-in region is not present.

## 4. Conclusions

The past 30 years have witnessed a great increase in our understanding of friction at small scales. Advances in instrumentation have led to the discovery of novel phenomena such as structural superlubricity and atomic scale stick-slip motion,[8], [44] while improvements in computational power have enabled detailed molecular dynamics simulations shedding new light on the physics of tribological interfaces.[45] Until now, however, there has been a dearth of experimental data on the frictional properties of plastically forming single asperity contacts, particularly in the very early stages of lateral loading. Connecting the seemingly disparate worlds of laboratory based small scale tribology with macroscopic friction, where asperity friction plays a prominent role, requires experiments such as ours that incorporate high normal pressures while retaining some of the high precision and accuracy seen in SFA or AFM studies.

The results of this paper indicate the following. The additional plastic deformation and sink-in caused by increased stress in the contact plays a critical role in setting the static coefficient of friction. The transition from static sink-in to kinetic ploughing is driven by failure of the interface and hence by adsorbates on the surfaces. It is also affected by the angle at the indent edge. We find that the E/H ratio (or the inverse, the yield strain ~ H/E) determines both the initial lateral deformation and the nature of the transition to full sliding. These results shed new light on the interaction between interface friction and bulk plasticity and may help to inform future studies linking friction at different scales.

## Acknowledgements

OB and GP gratefully acknowledge support from the Texas A&M University President's Excellence Fund X-Grants Program.

JP acknowledges discussions with the late David Tabor that stimulated this work.

**Bibliography**

[1]   F. P. Bowden and D. Tabor, *The friction and lubrication of solids*. Oxford: Oxford University Press, 1950.

[2]   I. Hutchings and P. Shipway, *Tribology - Friction and wear of engineering materials*, 2nd ed. Oxford: Butterworth-Heineman, 2017.

[3]   C. M. Mate and R. W. Carpick, *Tribology on the small scale : a modern textbook on friction, lubrication and wear.*, 2nd ed. Oxford: Oxford University Press, 2019.

[4]   P. J. Blau, "On the nature of running-in," *Tribol. Int.*, vol. 38, no. 11-12 SPEC. ISS., pp. 1007–1012, Nov. 2005.

[5]   D. Tabor, "Friction-the present state of our understanding," *J. Tribol.*, vol. 103, no. 2, pp. 169–179, Apr. 1981.

[6]   R. W. Carpick and M. Salmeron, "Scratching the surface: Fundamental investigations of tribology with atomic force microscopy," *Chem. Rev.*, vol. 97, no. 4, pp. 1163–1194, 1997.

[7]   I. Szlufarska, M. Chandross, and R. W. Carpick, "Recent advances in single-asperity nanotribology," *J. Phys. D. Appl. Phys.*, vol. 41, no. 12, Jun. 2008.

[8]   C. M. Mate, G. M. McClelland, R. Erlandsson, and S. Chiang, "Atomic-Scale Friction of a Tungsten Tip on a Graphite Surface," Springer, Dordrecht, 1987, pp. 226–229.

[9]   H. Gang, M. H. Müser, and M. O. Robbins, "Adsorbed layers and the origin of static friction," *Science (80-. ).*, vol. 284, no. 5420, pp. 1650–1652, Jun. 1999.

[10]  M. Mishra, P. Egberts, R. Bennewitz, and I. Szlufarska, "Friction model for single-

asperity elastic-plastic contacts," *Phys. Rev. B - Condens. Matter Mater. Phys.*, vol. 86, no. 4, p. 045452, Jul. 2012.

[11] A. Kareer, E. Tarleton, C. Hardie, S. V. Hainsworth, and A. J. Wilkinson, "Scratching the surface: Elastic rotations beneath nanoscratch and nanoindentation tests," *Acta Mater.*, vol. 200, pp. 116–126, Nov. 2020.

[12] J. R. Barber, "Multiscale surfaces and Amontons' Law of Friction," *Tribol. Lett.*, vol. 49, no. 3, pp. 539–543, Mar. 2013.

[13] K. Nakano and V. L. Popov, "Dynamic stiction without static friction: The role of friction vector rotation," *Phys. Rev. E*, vol. 102, no. 6, p. 063001, Dec. 2020.

[14] D. F. Ogletree, R. W. Carpick, and M. Salmeron, "Calibration of frictional forces in atomic force microscopy," *Rev. Sci. Instrum.*, vol. 67, no. 9, pp. 3298–3306, Sep. 1996.

[15] S. Lafaye and M. Troyon, "On the friction behaviour in nanoscratch testing," *Wear*, 2006.

[16] S. Lafaye, C. Gauthier, and R. Schirrer, "Analyzing friction and scratch tests without in situ observation," *Wear*, vol. 265, no. 5–6, pp. 664–673, Aug. 2008.

[17] A. M. Homola, J. N. Israelachvili, P. M. McGuiggan, and M. L. Gee, "Fundamental experimental studies in tribology: The transition from 'interfacial' friction of undamaged molecularly smooth surfaces to 'normal' friction with wear," *Wear*, vol. 136, no. 1, pp. 65–83, Feb. 1990.

[18] B. N. J. Persson, O. Albohr, F. Mancosu, V. Peveri, V. N. Samoilov, and I. M. Sivebaek, "On the nature of the static friction, kinetic friction and creep," *Wear*, vol. 254, no. 9, pp. 835–851, May 2003.

[19]  Q. Li, T. E. Tullis, D. Goldsby, and R. W. Carpick, "Frictional ageing from interfacial bonding and the origins of rate and state friction," *Nature*, vol. 480, no. 7376, pp. 233–236, Dec. 2011.

[20]  Y. Liu and I. Szlufarska, "Chemical origins of frictional aging," *Phys. Rev. Lett.*, vol. 109, no. 18, p. 186102, Nov. 2012.

[21]  J. H. Dieterich and B. D. Kilgore, "Direct observation of frictional contacts: New insights for state-dependent properties," *Pure Appl. Geophys. PAGEOPH*, vol. 143, no. 1–3, pp. 283–302, Mar. 1994.

[22]  D. L. Goldsby, A. Rar, G. M. Pharr, and T. E. Tullis, "Nanoindentation creep of quartz, with implications for rate- and state-variable friction laws relevant to earthquake mechanics," *J. Mater. Res.*, vol. 19, no. 1, pp. 357–365, Jan. 2004.

[23]  O. Brazil and G. M. Pharr, "Direct observation of partial interface slip in micrometre-scale single asperity contacts," *Tribol. Int.*, p. 106776, Nov. 2020.

[24]  D. Tabor, "Junction growth in metallic friction: the role of combined stresses and surface contamination," *Proc. R. Soc. London. Ser. A. Math. Phys. Sci.*, vol. 251, no. 1266, pp. 378–393, Jun. 1959.

[25]  J. A. Nieminen, A. P. Sutton, and J. B. Pethica, "Static junction growth during frictional sliding of metals," *Acta Metall. Mater.*, vol. 40, no. 10, pp. 2503–2509, Oct. 1992.

[26]  K. L. Johnson, *Contact Mechanics*. Cambridge: Cambridge University Press, 1985.

[27]  J. M. Challen, L. J. McLean, and P. L. B. Oxley, "Plastic deformation of a metal surface in sliding contact with a hard wedge: its relation to friction and wear," *Proc. R. Soc. London. A. Math. Phys. Sci.*, vol. 394, no. 1806, pp. 161–181, Jul. 1984.

[28] J. B. Pethica and W. C. Oliver, "Mechanical Properties of Nanometre Volumes of Material: use of the Elastic Response of Small Area Indentations," *MRS Proc.*, vol. 130, 1988.

[29] J. Hay, P. Agee, and E. Herbert, "Continuous stiffness measurement during instrumented indentation testing," *Exp. Tech.*, vol. 34, no. 3, pp. 86–94, May 2010.

[30] G. M. Pharr, J. H. Strader, and W. C. Oliver, "Critical issues in making small-depth mechanical property measurements by nanoindentation with continuous stiffness measurement," *J. Mater. Res.*, vol. 24, no. 3, pp. 653–666, Mar. 2009.

[31] B. Merle, V. Maier-Kiener, and G. M. Pharr, "Influence of modulus-to-hardness ratio and harmonic parameters on continuous stiffness measurement during nanoindentation," *Acta Mater.*, vol. 134, pp. 167–176, Aug. 2017.

[32] G. M. Pharr, W. C. Oliver, and F. R. Brotzen, "On the generality of the relationship among contact stiffness, contact area, and elastic modulus during indentation," *J. Mater. Res.*, vol. 7, no. 3, pp. 613–617, 1992.

[33] F. P. Bowden and T. H. . Childs, "The friction and deformation of clean metals at very low temperatures," *Proc. R. Soc. London. A. Math. Phys. Sci.*, vol. 312, no. 1511, pp. 451–466, Sep. 1969.

[34] J. S. Courtney-Pratt and E. Eisner, "The effect of a tangential force on the contact of metallic bodies," *Proc. R. Soc. London. Ser. A. Math. Phys. Sci.*, vol. 238, no. 1215, pp. 529–550, Jan. 1957.

[35] F. P. Bowden and J. E. Young, "Friction of clean metals and the influence of adsorbed films," *Proc. R. Soc. London. Ser. A. Math. Phys. Sci.*, vol. 208, no. 1094, pp. 311–325, Sep. 1951.

[36] T. Kayaba and K. Kato, "Experimental analysis of junction growth with a junction model," *Wear*, vol. 51, no. 1, pp. 105–116, Nov. 1978.

[37] A. Ovcharenko, G. Halperin, and I. Etsion, "In situ and real-time optical investigation of junction growth in spherical elastic-plastic contact," *Wear*, vol. 264, no. 11–12, pp. 1043–1050, May 2008.

[38] J. L. Bucaille, E. Felder, and G. Hochstetter, "Mechanical analysis of the scratch test on elastic perfectly plastic materials with the three-dimensional finite element modeling," *Wear*, vol. 249, no. 5–6, pp. 422–432, Jun. 2001.

[39] B. D. Beake, A. J. Harris, and T. W. Liskiewicz, "Review of recent progress in nanoscratch testing," *Tribol. - Mater. Surfaces Interfaces*, vol. 7, no. 2, pp. 87–96, Jun. 2013.

[40] A. Kareer, X. D. Hou, N. M. Jennett, and S. V. Hainsworth, "The existence of a lateral size effect and the relationship between indentation and scratch hardness in copper," *Philos. Mag.*, vol. 96, no. 32–34, pp. 3396–3413, Dec. 2016.

[41] S. Li *et al.*, "The evolving quality of frictional contact with graphene," *Nature*, vol. 539, no. 7630, pp. 541–545, Nov. 2016.

[42] J. B. Pethica and D. Tabor, "Contact of characterised metal surfaces at very low loads: Deformation and adhesion," *Surf. Sci.*, vol. 89, no. 1–3, pp. 182–190, Jan. 1979.

[43] P. M. Hoffmann, S. Jeffery, J. B. Pethica, H. Özgür Özer, and A. Oral, "Energy dissipation in atomic force microscopy and atomic loss processes," *Phys. Rev. Lett.*, vol. 87, no. 26, pp. 265502-1-265502–4, Dec. 2001.

[44] M. Dienwiebel, G. S. Verhoeven, N. Pradeep, J. W. M. Frenken, J. A. Heimberg, and H. W. Zandbergen, "Superlubricity of graphite," *Phys. Rev. Lett.*, vol. 92, no. 12, p.

126101, Mar. 2004.

[45] A. Vanossi *et al.*, "Recent highlights in nanoscale and mesoscale friction," *Beilstein J. Nanotechnol.*, vol. 9, no. 1, pp. 1995–2014, 2018.